\documentclass{article}
\usepackage{epsfig}
\usepackage[latin1]{inputenc}
\DeclareFontFamily{OT1}{bbold}{}

\DeclareFontShape{OT1}{bbold}{m}{n}{<5><6><7><8><9> sgen * bbold
                                    <10-12> bbold10
                                    <12-14> bbold12
                                    <14-17> bbold14
                                    <17-> bbold17}{}
\DeclareMathAlphabet{\mathbb}{OT1}{bbold}{m}{n}
\newcommand{\IM}{[\mathbb{1}]}
\usepackage{xspace}
\usepackage{a4wide}
\newcommand{\eps}{\varepsilon}

\def\S12{\mathrm{S}_{1,2}}

\newcommand{\halb}{\mbox{$\frac{1}{2}$}}

\newcommand{\viertel}{\mbox{$\frac{1}{4}$}}

\def\slash#1{\setbox0=\hbox{$\displaystyle#1$}
        \setbox1=\hbox{$\scriptstyle#1$}\setbox2=\hbox{$\scriptscriptstyle#1$}
        \dimen0=\wd0 \dimen1=\wd1 \dimen2=\wd2
        \divide\dimen0 by 2 \advance\dimen0 by 0.6ex
        \divide\dimen1 by 2 \advance\dimen1 by 0.45ex
        \divide\dimen2 by 2 \advance\dimen2 by 0.35ex
        \ifx #1c \advance\dimen0 by -0.1ex \fi
        \ifx #1e \advance\dimen0 by -0.1ex \fi
        \ifx #1g \advance\dimen0 by -0.1ex \fi
        \ifx #1o \advance\dimen0 by -0.1ex \fi
        \ifx #1p \advance\dimen0 by -0.1ex \fi
        \ifx #1s \advance\dimen0 by -0.1ex \fi
        \ifx #1v \advance\dimen0 by -0.1ex \fi
        \ifx #1y \advance\dimen0 by -0.1ex \fi
        \ifx #1z \advance\dimen0 by -0.1ex \fi
        \ifx #1A \advance\dimen0 by -0.1ex \fi
        \ifx #1U \advance\dimen0 by 0.1ex \fi
        \ifx #1V \advance\dimen0 by 0.1ex \fi
        \ifx #1W \advance\dimen0 by 0.1ex \fi
        \ifx #1Y \advance\dimen0 by 0.1ex \fi
        \setbox3=\hbox{$\displaystyle#1\hskip-\dimen0%
        /\hskip\dimen0\hskip-1.2ex$}
        \setbox4=\hbox{$\textstyle#1\hskip-\dimen0%
        /\hskip\dimen0\hskip-1.2ex$}
        \setbox5=\hbox{$\scriptstyle#1\hskip-\dimen1%
        /\hskip\dimen1\hskip-1.2ex$}
        \setbox6=\hbox{$\scriptscriptstyle#1\hskip-\dimen2%
        /\hskip\dimen2\hskip-1.2ex$}
        \mathchoice{\box3}{\box4}{\box5}{\box6}}

\newcounter{allequation}
\newcommand{\subnumbers}{
     \setcounter{allequation}{\value{equation}}
      \addtocounter{allequation}{1}
      \setcounter{equation}{0}
       \renewcommand{\theequation}{\arabic{allequation}\alph{equation}}}
\newcommand{\normalnumbers}{\setcounter{equation}{\value{allequation}}
                       \renewcommand{\theequation}{\arabic{equation}}\\}

\begin{document}
\title{
  {Renormalization of the Cabibbo-Kobayashi-Maskawa Matrix}\\[1cm]} 
\author{A. Barroso${}^{a,b,1}$, L.
  Brücher${}^{b,2}$ and 
  R. Santos${}^{b,c,3}$ \\[0.5cm]
{\small \em ${}^{a}$Dept. de F\'\i sica, Faculdade de Ci\^encias, Universidade de Lisboa}\\
{\small \em Campo Grande, C1, 1749-016 Lisboa, Portugal}\\[0.5cm]
{\small \em ${}^{b}$Universidade de Lisboa, Centro de F\'\i sica Nuclear,}\\
{\small \em  Avenida Professor Gama Pinto 2, 1649-003 Lisboa, Portugal}\\[0.5cm] 
{\small \em ${}^{c}$Instituto Superior de Transportes, Campus Universit\'ario,}\\
{\small \em Rua D. Afonso Henriques, 2330 Entroncamento, Portugal}\\[2.5cm]}
\date{} 

\maketitle

\begin{abstract}
Using the on-shell scheme and the general linear $R_\xi$ gauge we have
calculated the one-loop amplitude $W^+ \rightarrow u_I \bar{d}_j$. In
agreement with previous work we have shown that the
Cabibbo-Kobayashi-Maskawa (CKM) matrix ought to be renormalized. We
show how to renormalize the CKM matrix and, at the same time, obtain a
gauge independent $W$ decay amplitude. 
\end{abstract}
\vspace*{3cm}

\begin{flushleft}
  PACS number(s): 11.10.Gh, 12.15.-y, 12.15.Ff, 12.15.Lk
\end{flushleft}

\footnotetext[1]{e-mail: barroso@alf1.cii.fc.ul.pt}
\footnotetext[2]{e-mail: bruecher@alf1.cii.fc.ul.pt}
\footnotetext[3]{e-mail: rsantos@alf1.cii.fc.ul.pt}

\thispagestyle{empty}

\newpage




\section{Introduction}

The electroweak sector of the standard model (SM) has been the subject
of extensive studies during the last twenty-five years. Since the
renormalizability of the SM was proved \cite{Ho8} an immense effort has
been made to implement this renormalization program at one-loop level
(cf. ref. \cite{Aok} and \cite{Bohm1} for a review). The agreement between
these calculations and the experimental results is impressive. 

Despite these facts, the renormalization of the
Cabibbo-Kobayashi-Maskawa (CKM) quark mixing matrix \cite{Cabi} was done
only by one group, Denner and Sack \cite{Den1} (DS) in 1990. They have
shown that, as soon as one takes into account the non-degeneracy of
the quark masses, the CKM ought to be renormalized. However, recently
Gambino, Grassi and Madricardo \cite{Gamb} (GGM) have raised some doubts about
the DS renormalization prescription. In particular, they have claimed
that the on-shell conditions used by DS 
lead to a gauge dependent width for the decay $W \rightarrow
q\bar{q}$. Then, they propose an alternative renormalization
prescription.

In view of this situation, we decided that it is appropriate to carry
out another independent calculation of the renormalization of the
CKM. This is our aim. We repeat the work of DS, but with a fundamental
difference. Rather than using the common 't Hooft Feynman gauge
($\xi=1$) we do our calculation in the general linear $R_\xi$
gauge. Hence, we will be able to show, explicitly, the problem raised
by GGM and make a proposal to solve it.

To address the question of the CKM renormalization one has to consider
a process where this matrix appears at tree-level. To be precise, let
us consider the decay $W^+ \rightarrow q_I \bar{q}_j$, where $I$ and
$j$ are generation indices. We use capital letters for the {\em
  up}-type quarks and lower case letters for the {\em down}-type
quarks. Then, at tree-level the decay amplitude $T_0$ is
\begin{equation}
  \label{eq1}
  T_0 = V_{Ij} A_L
\end{equation}
with
\begin{equation}
  \label{eq2}
  A_L = \frac{g\,N_c}{\sqrt{2}}\,\bar{u}_I(p_1)\slash{\eps}\gamma_L
  v_j(q-p_1) \enskip .
\end{equation}
$V_{Ij}$ are the elements of the CKM matrix, $N_c$ is the number of
colors and $g$ is the $SU(2)$ coupling constant.

At one-loop eq.~(\ref{eq1}) is modified in several different
ways. Firstly, one has to sum all one-loop irreducible
vertex-diagrams. This gives a contribution proportional to $V_{Ij}$
but not entirely proportional to $A_L$. Secondly, we have the
counter terms stemming from the usual variation of the Lagrangian
parameters. The counter terms $\delta g$ and $\delta Z_W$ ($W$-wave
function renormalization) also give rise to contributions proportional
to the tree-level amplitude. However, since the quarks get mixed by
the renormalization procedure, this is not true for the quark wave
function renormalization constants $\delta Z_{II^\prime}^L$ and $\delta
Z_{jj^\prime}^L$. Finally, an additional counter term $\delta V_{Ij}$
has to be included. 

For a real $W$ that decays into on-shell quarks, it is easy to show that the
vertex diagrams can be written in terms of four independent form
factors. Each one is associated with a given Lorentz structure for the
spinors. Denoting by $q^\mu$ the 4-momentum of the incoming $W^+$ and
by $p_1^\mu$ the 4-momentum of the outgoing {\em up}-quark $I$, let us
define 
\begin{equation}
  \label{eq3}
  B_L = \frac{g\,N_c}{\sqrt{2}}\,\bar{u}_I(p_1)
  \frac{\eps\cdot p_1}{m_W} \gamma_L v_j(q-p_1) \enskip ,
\end{equation}
where $\eps^\mu$ is the $W$ polarization vector. Similarly, replacing
in eqs.~(\ref{eq2}) and (\ref{eq3}) $\gamma_L$ by $\gamma_R$ we define
$A_R$ and $B_R$ respectively. Now, the one-loop amplitude $T_1$ is
\begin{eqnarray}
  \label{eq4}
  T_1 & = & A_L \left [\, V_{Ij}\,\left(\,F_L +\frac{\delta g}{g} + \halb
  \delta Z_W + \halb \delta Z_{II}^{L*} + \halb \delta Z_{jj}^L\right)
   \,+\,\sum_{I^\prime \neq I} \halb \delta Z_{I^\prime I}^{L*}
  V_{I^\prime j} 
  \right. \nonumber \\ & & \left. \phantom{A_L [}
  \,+\,\sum_{j^\prime \neq j} V_{Ij^\prime} \halb
  \delta Z_{j^\prime j}^{L} \,+\, \delta V_{Ij} \,\right] 
  \,+\, V_{Ij} \, \left [ \, A_R F_R + B_L G_L + B_R G_R \,\right]
  \enskip ,
\end{eqnarray}
where $F_{L,R}$ and $G_{L,R}$ are the form-factors. We calculate the
different terms in eq.~(\ref{eq4}) using the general $R_\xi$ gauge for
the $W$-propagators. However to simplify the calculation, we use
't Hooft-Feynman gauge for the $Z$ and photon propagators. This is not
inconsistent, since the $\xi$ parameters of the gauge fixing
Lagrangian,
\begin{displaymath}
  {\cal L}_{GF} = -\frac{1}{2\xi_\gamma}(\partial\!\cdot\! A)^2 -
  \frac{1}{2\xi_Z}\left(\partial\!\cdot\! Z -\xi_Z m_Z G^0\right)^2
  -\frac{1}{\xi_W}\left| \partial\!\cdot\! W^+ + i\xi_W m_W G^+ \right|^2
\end{displaymath}
are independent. For our purpose it is sufficient to set $\xi_\gamma =
\xi_Z =1$ but to keep $\xi_W$ as a free parameter. From this point
onwards it will be denoted simply by $\xi$. For the
numerical calculations we used the values from Particle Data Group
\cite{Caso}.

\section{The irreducible vertex diagrams}

\begin{figure}[htbp]
  \begin{center}
    \epsfig{file=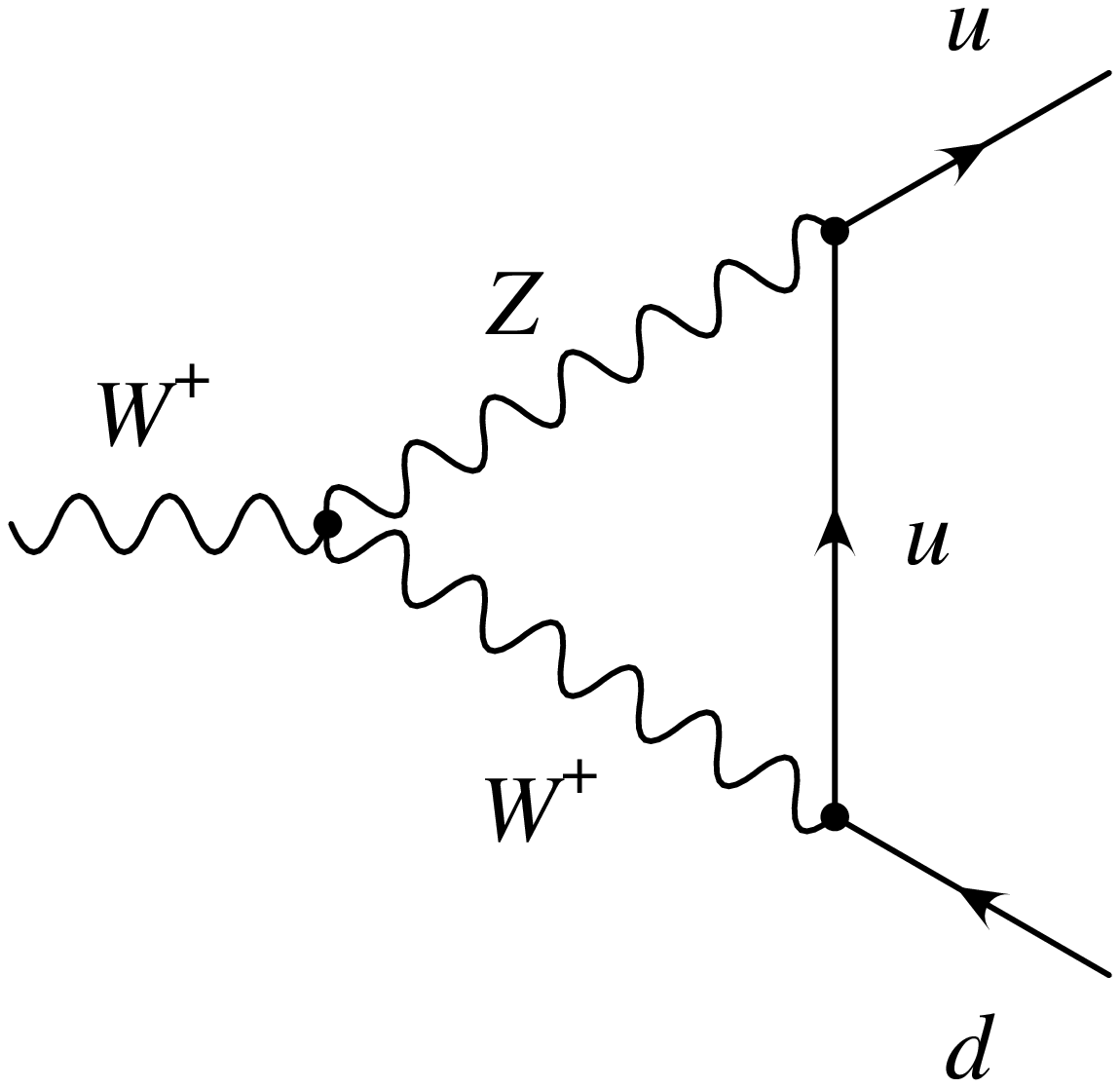,width=3.6cm}
    \epsfig{file=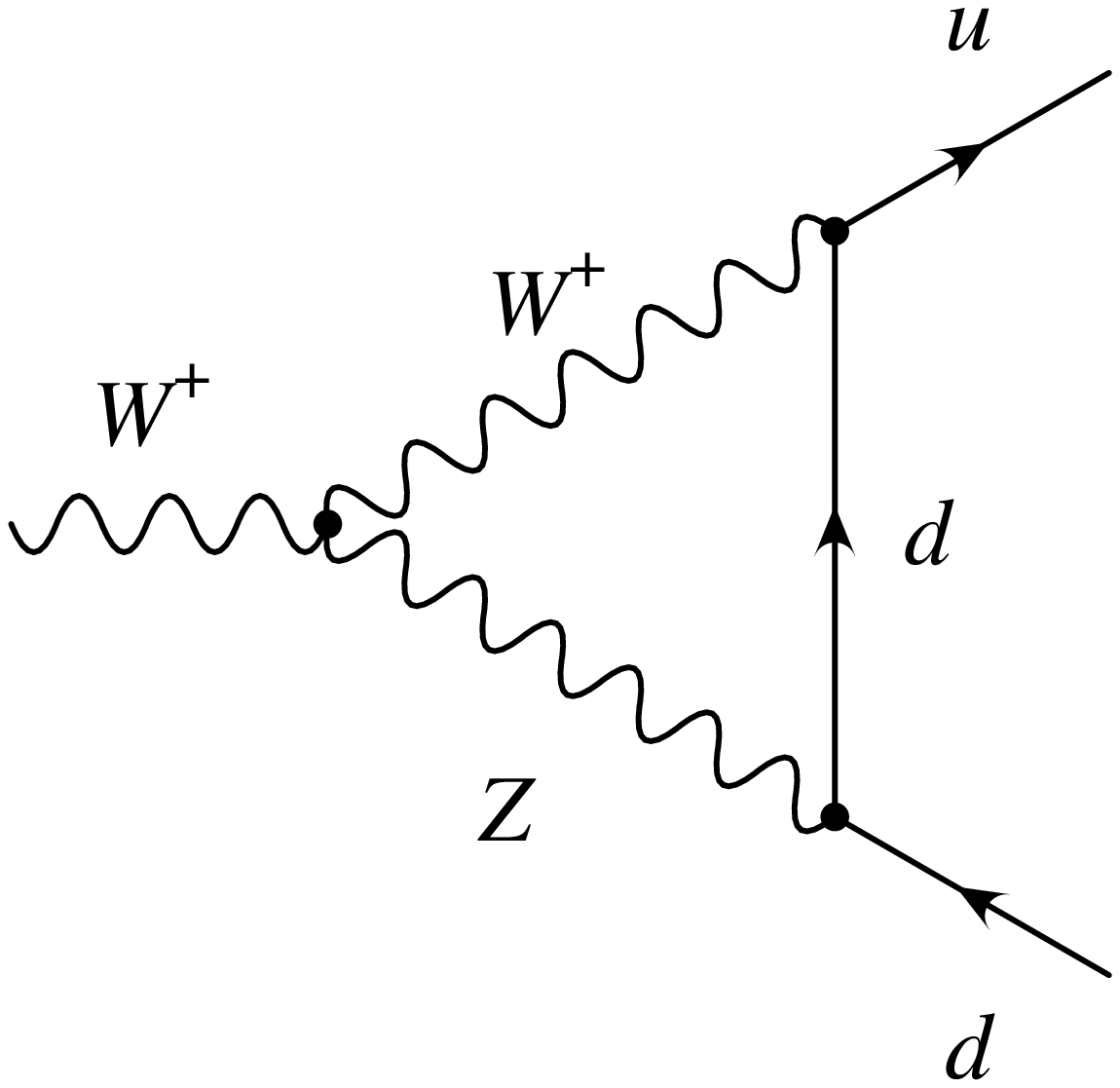,width=3.6cm}
    \epsfig{file=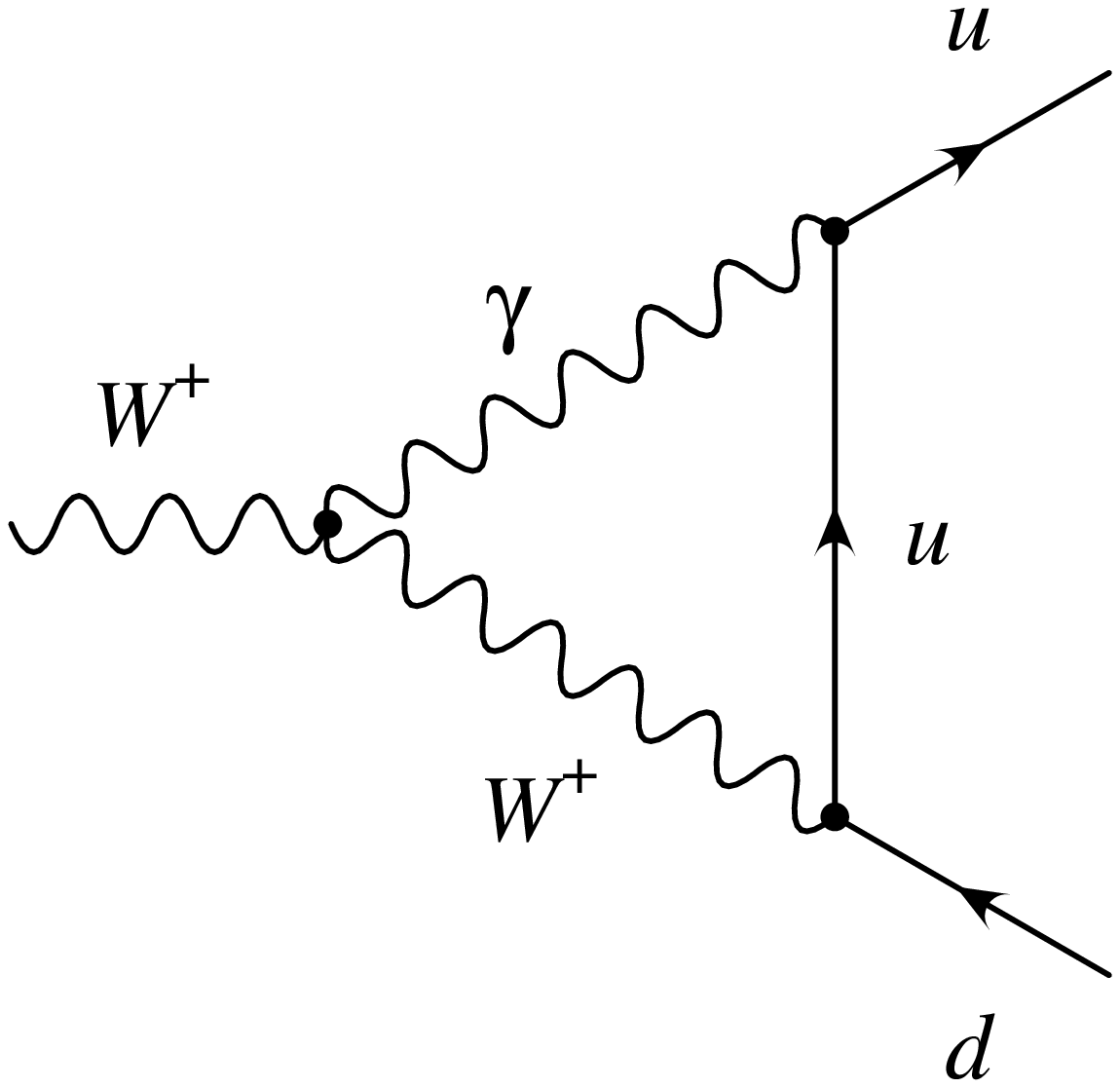,width=3.6cm}
    \epsfig{file=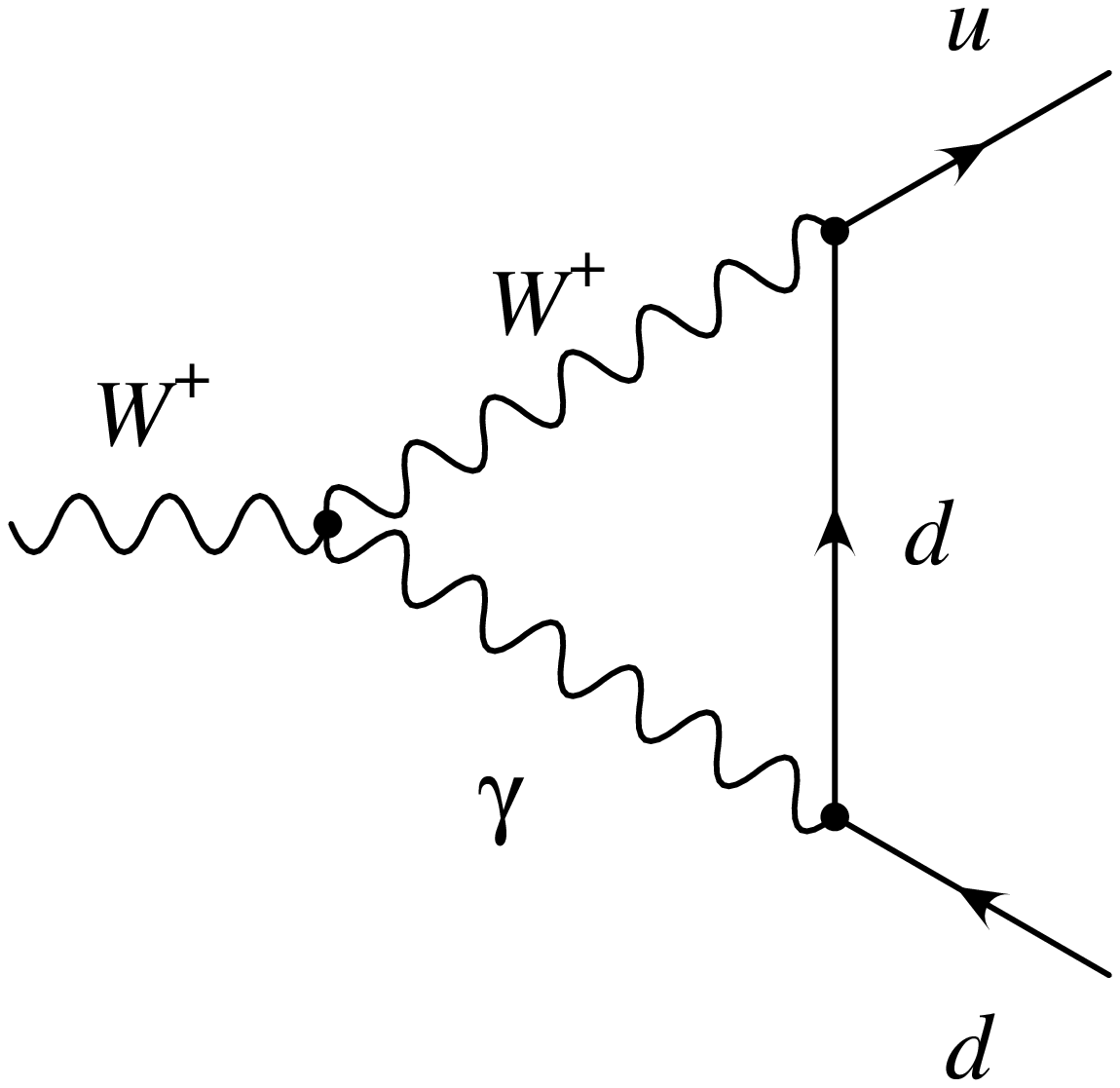,width=3.6cm}\\
    \epsfig{file=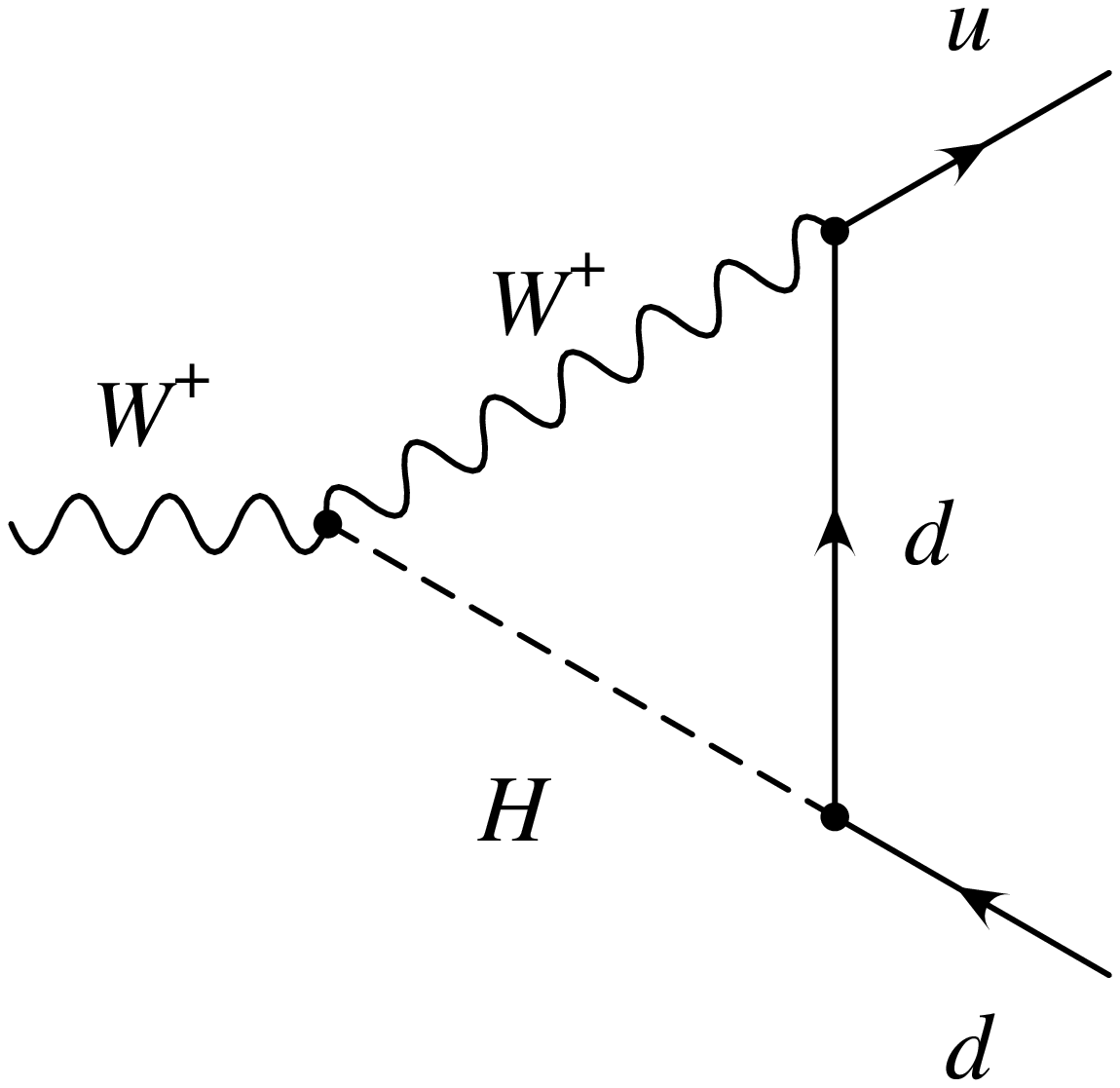,width=3.6cm}
    \epsfig{file=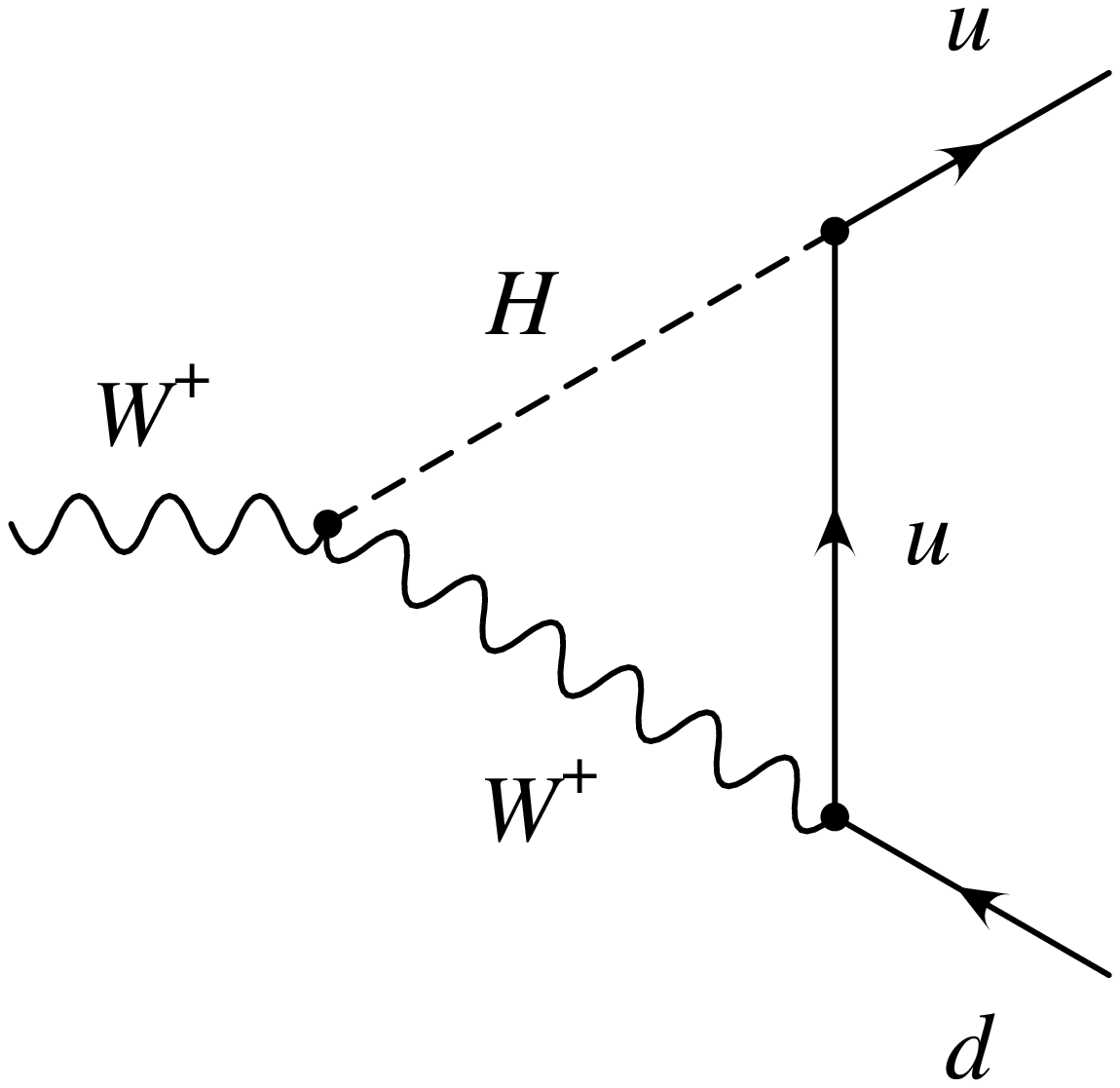,width=3.6cm}
    \epsfig{file=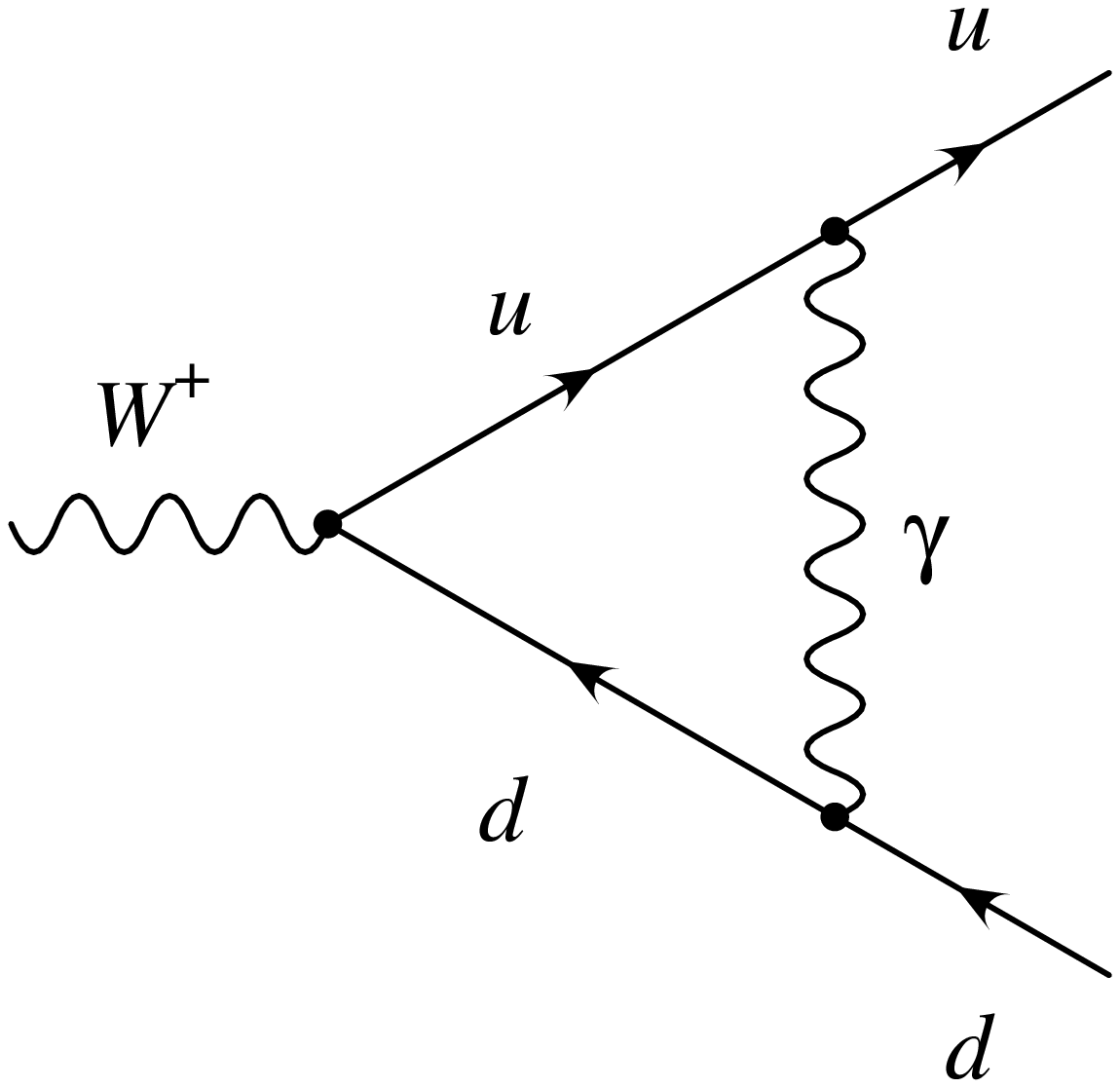,width=3.6cm}
    \epsfig{file=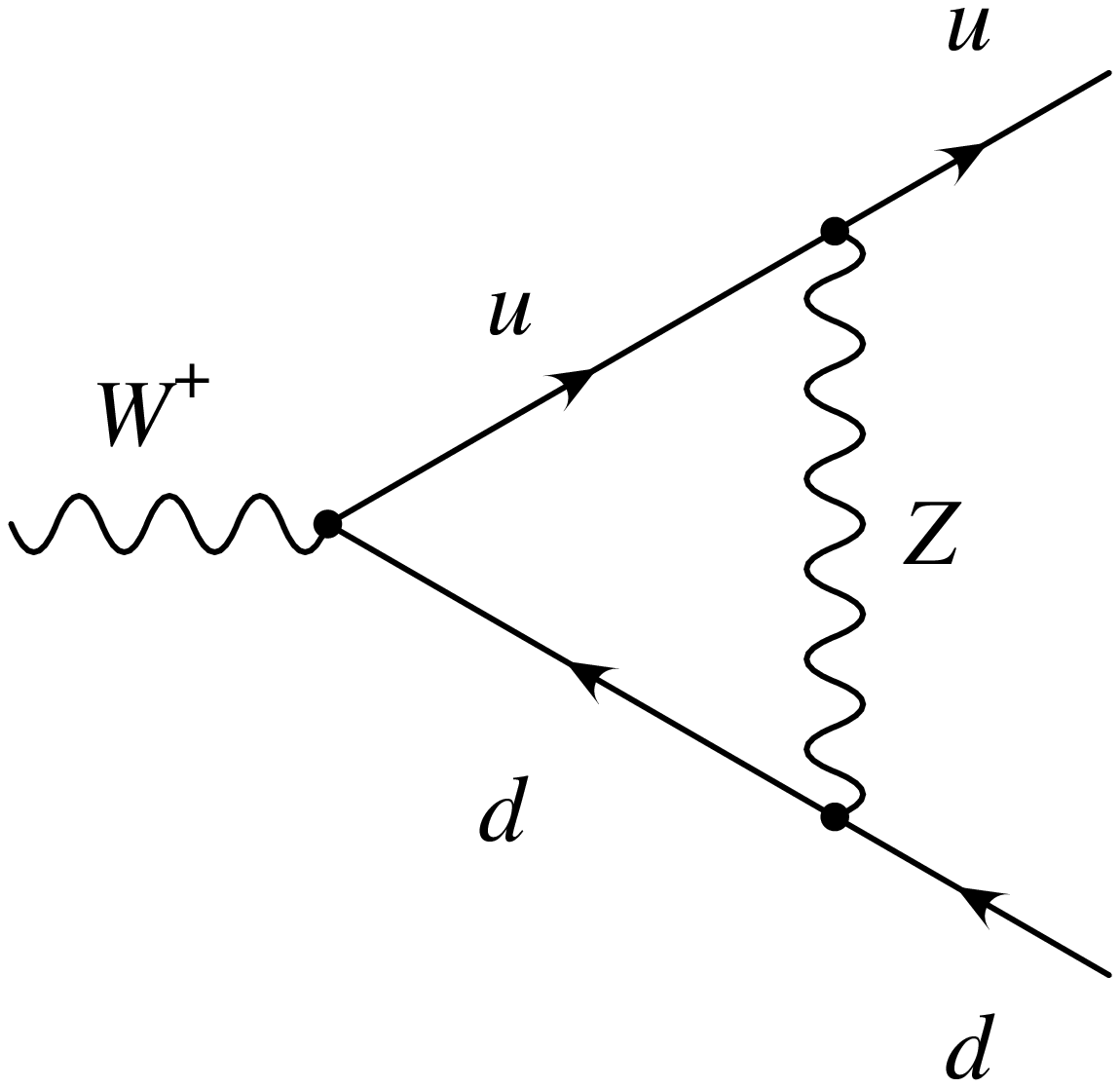,width=3.6cm}\\
    \epsfig{file=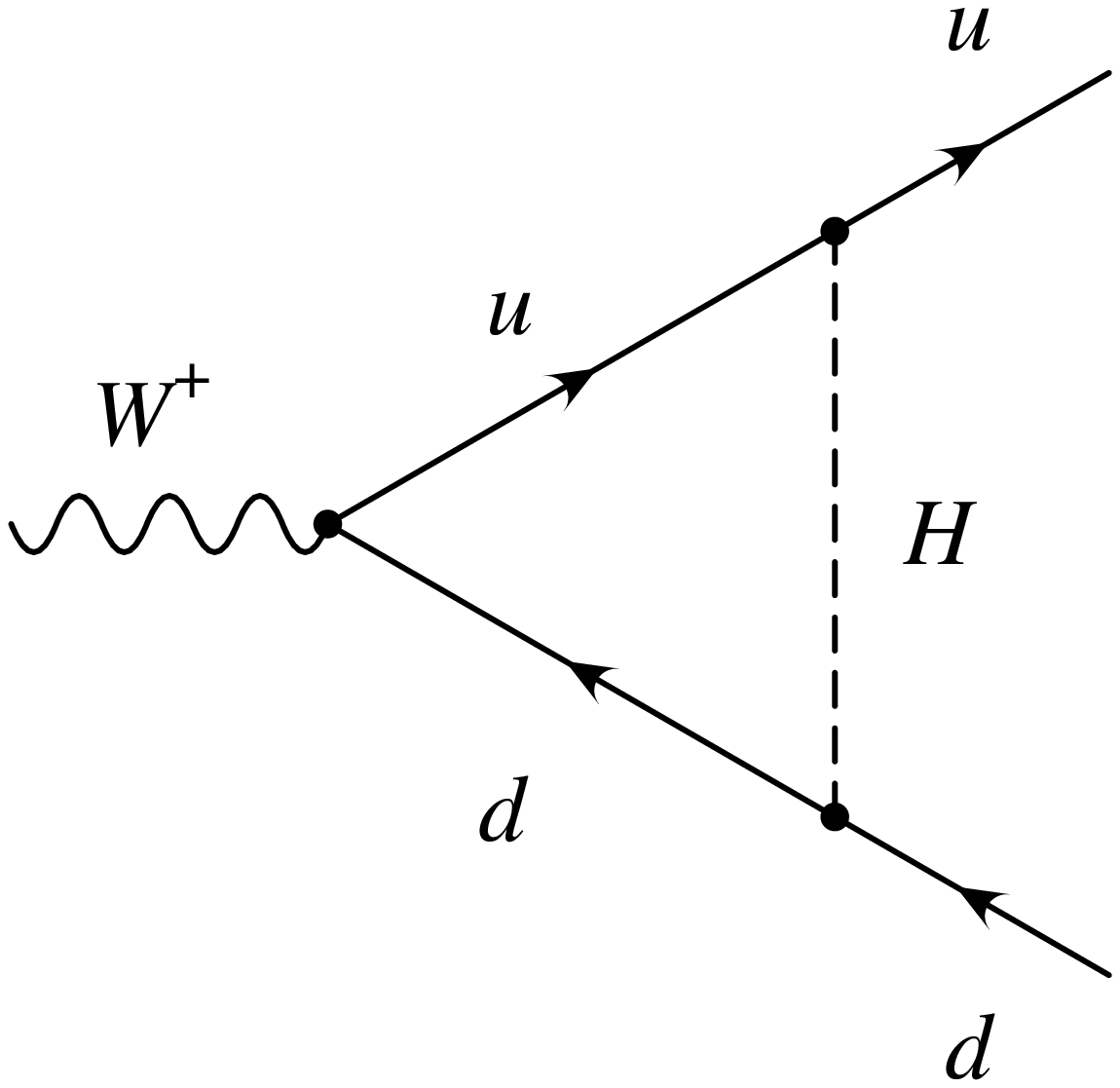,width=3.6cm}
    \epsfig{file=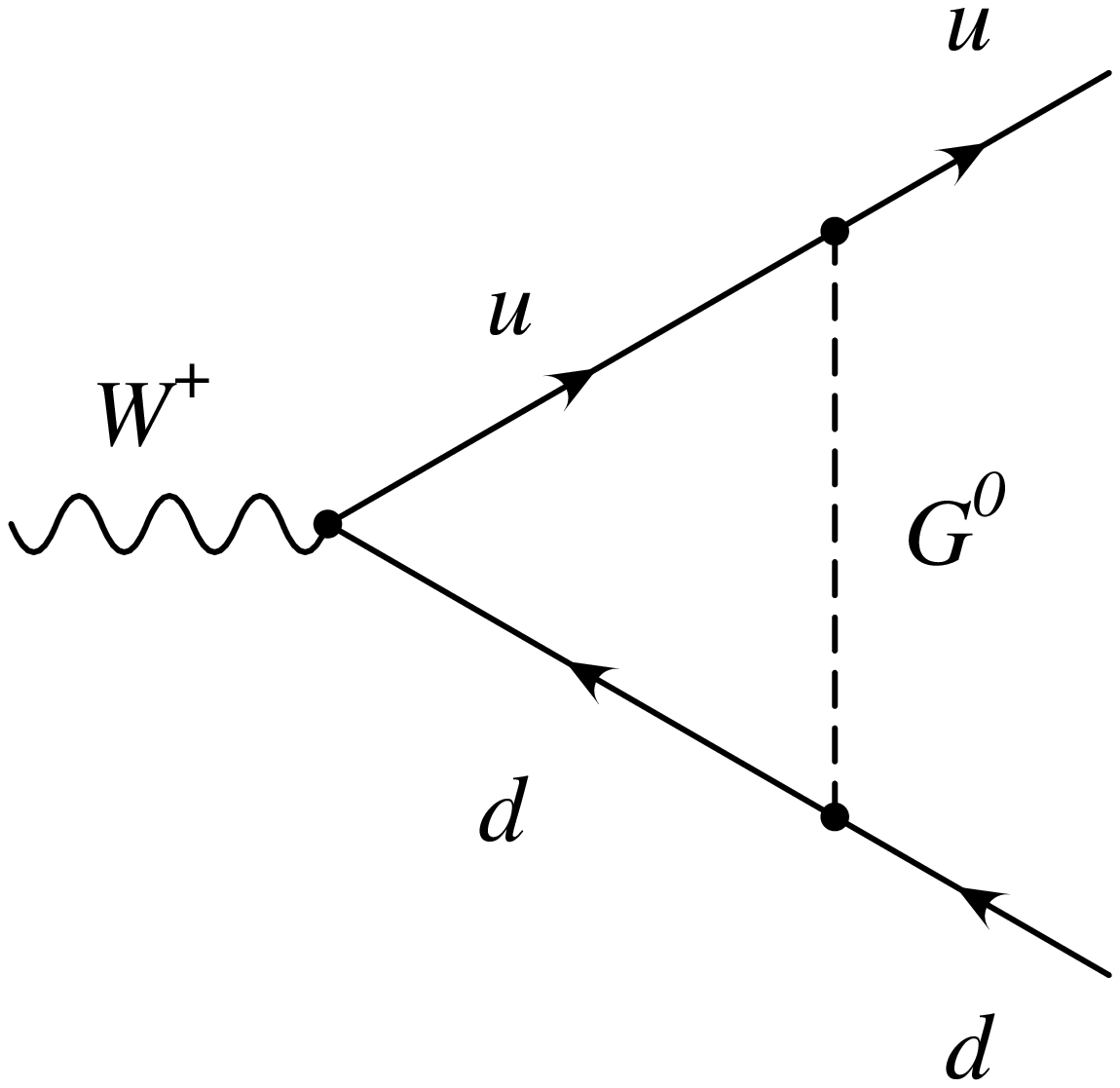,width=3.6cm}
    \epsfig{file=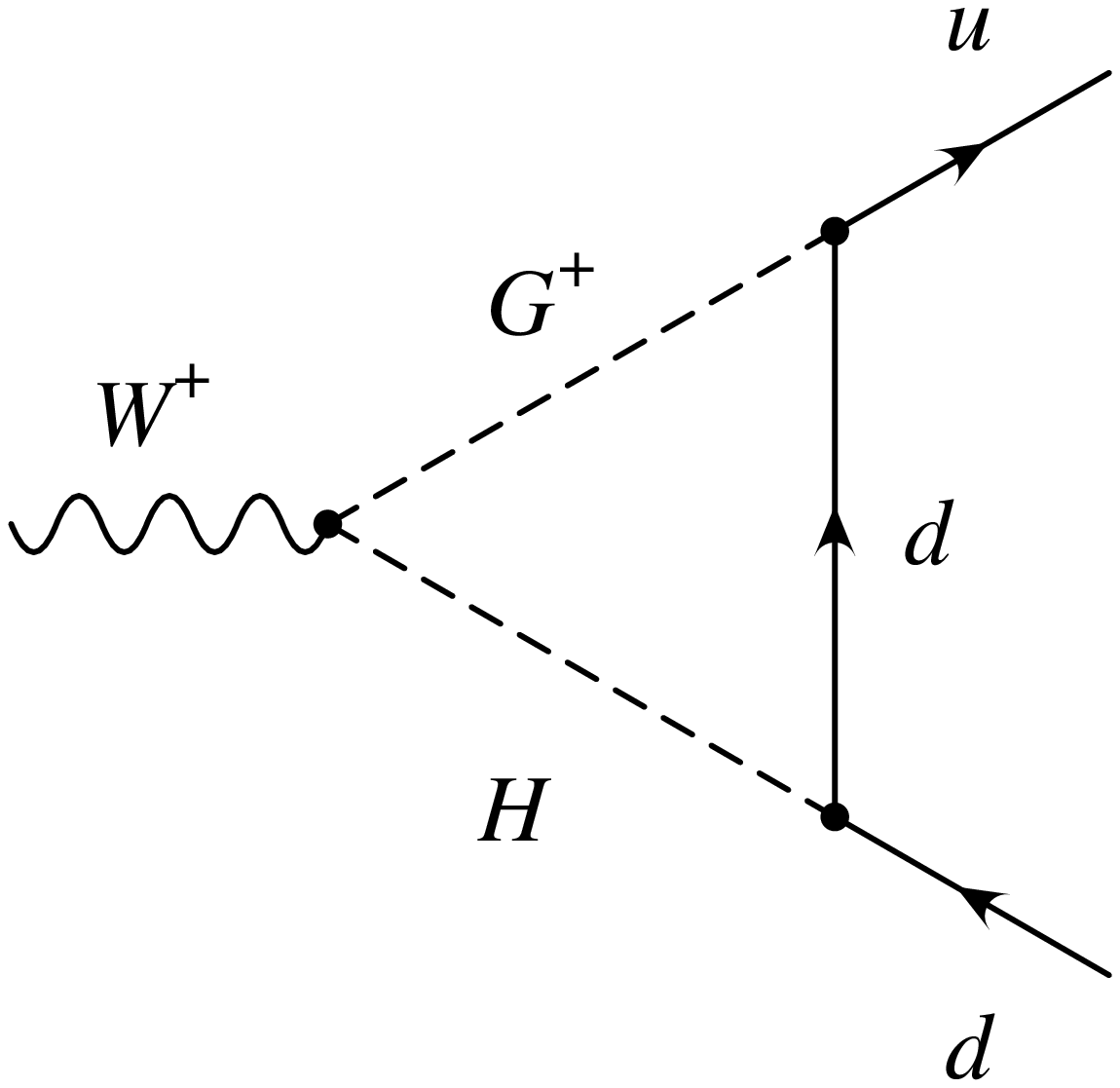,width=3.6cm}
    \epsfig{file=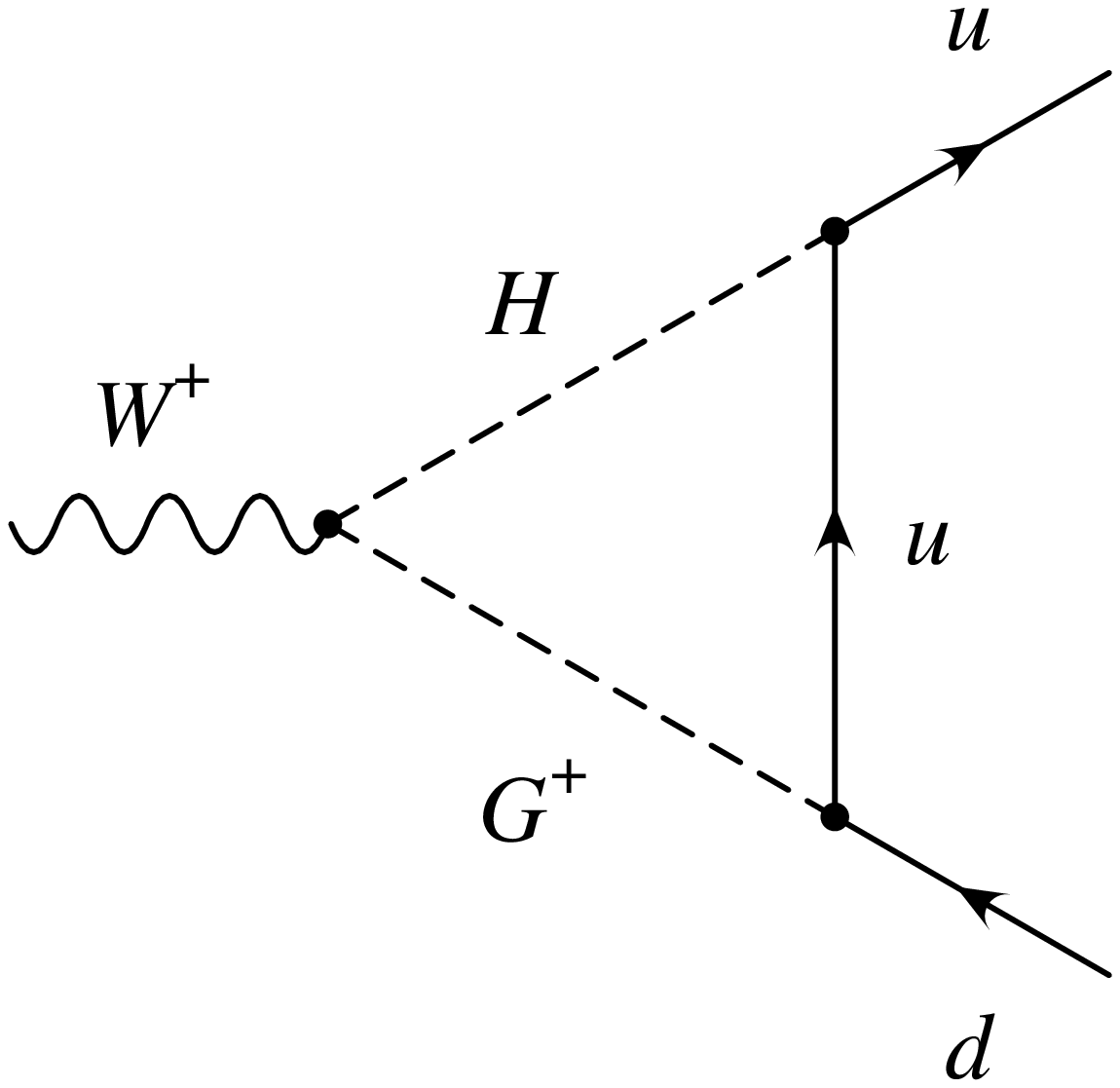,width=3.6cm}\\
    \epsfig{file=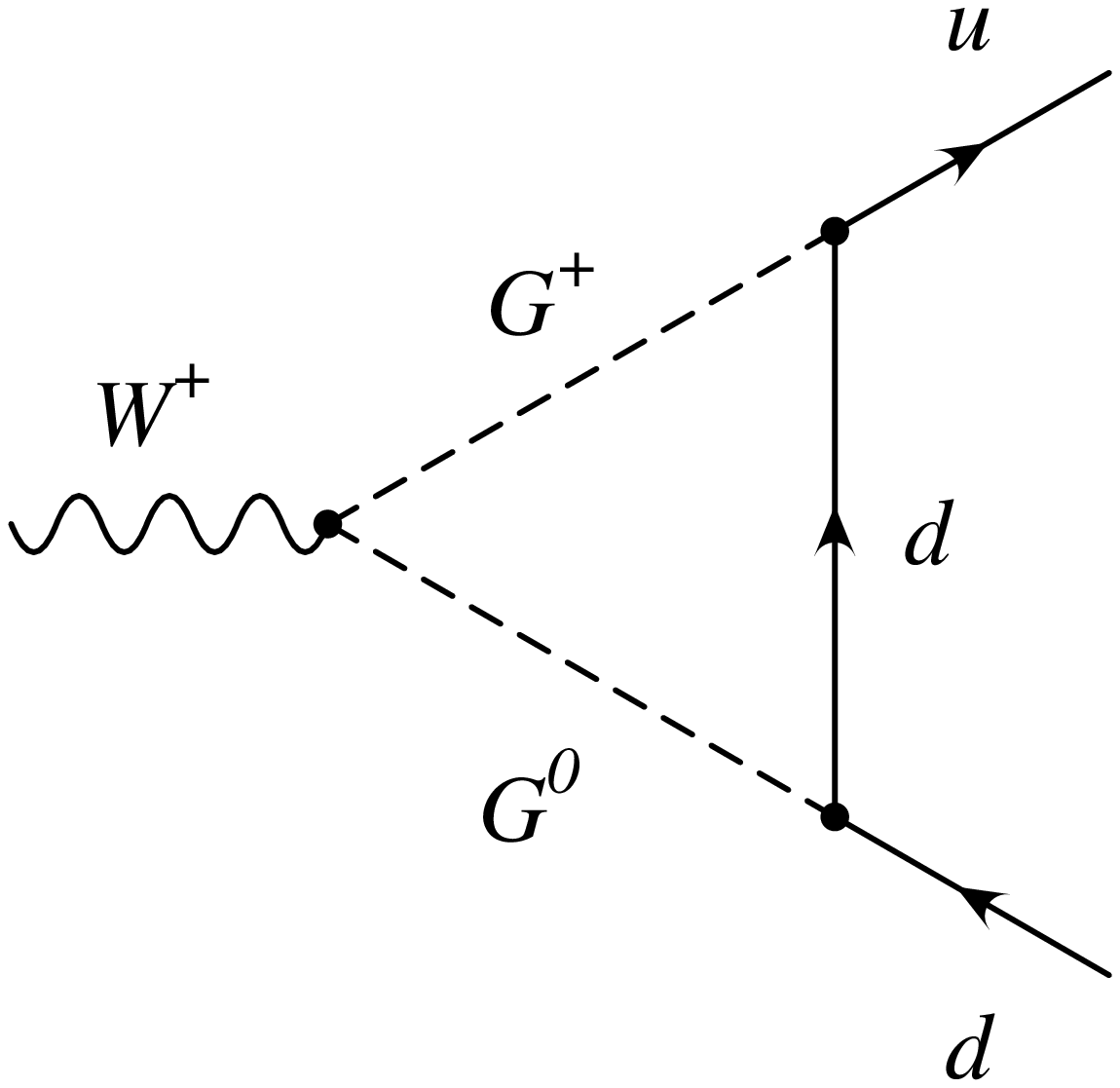,width=3.6cm}
    \epsfig{file=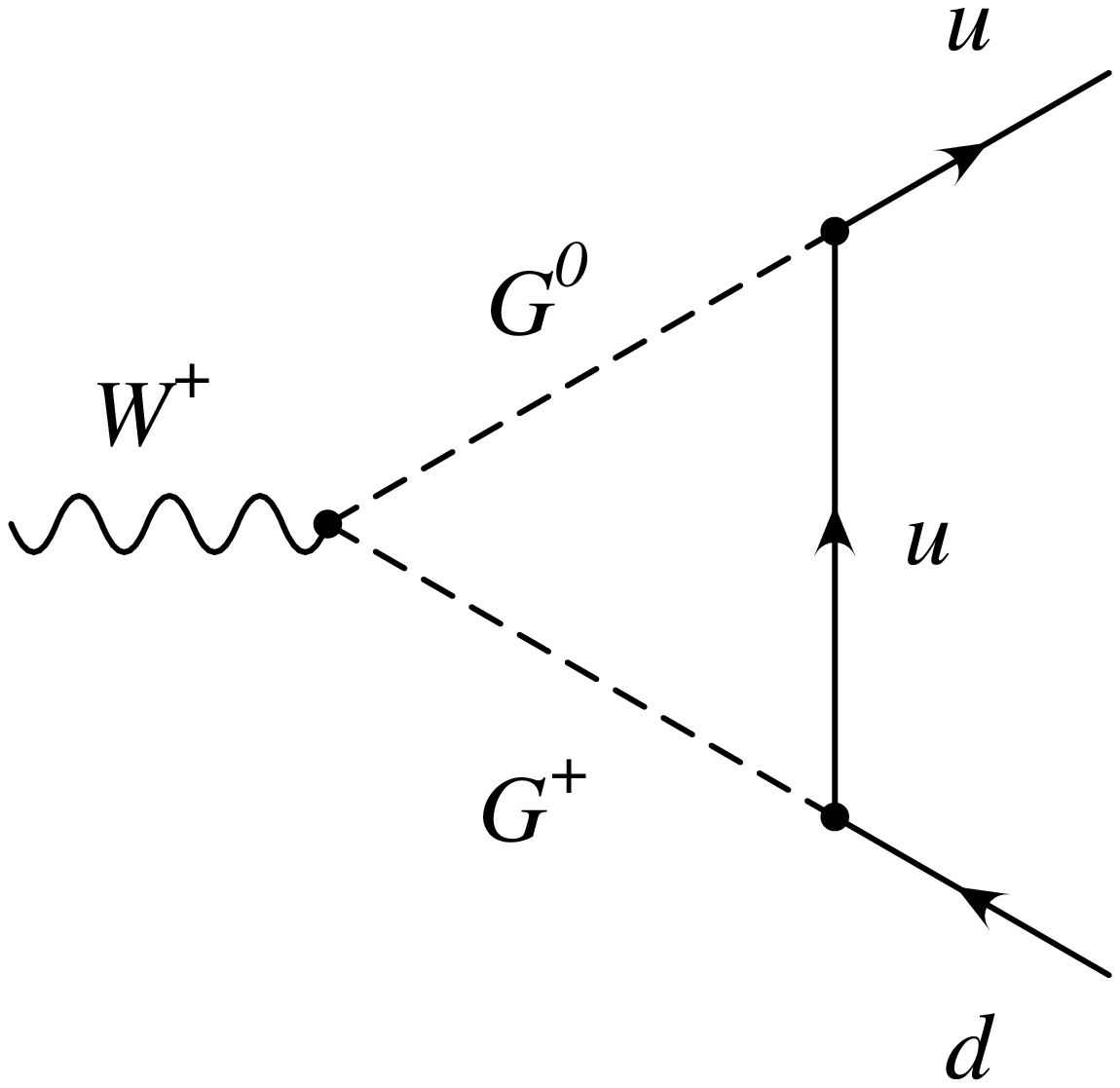,width=3.6cm}
    \epsfig{file=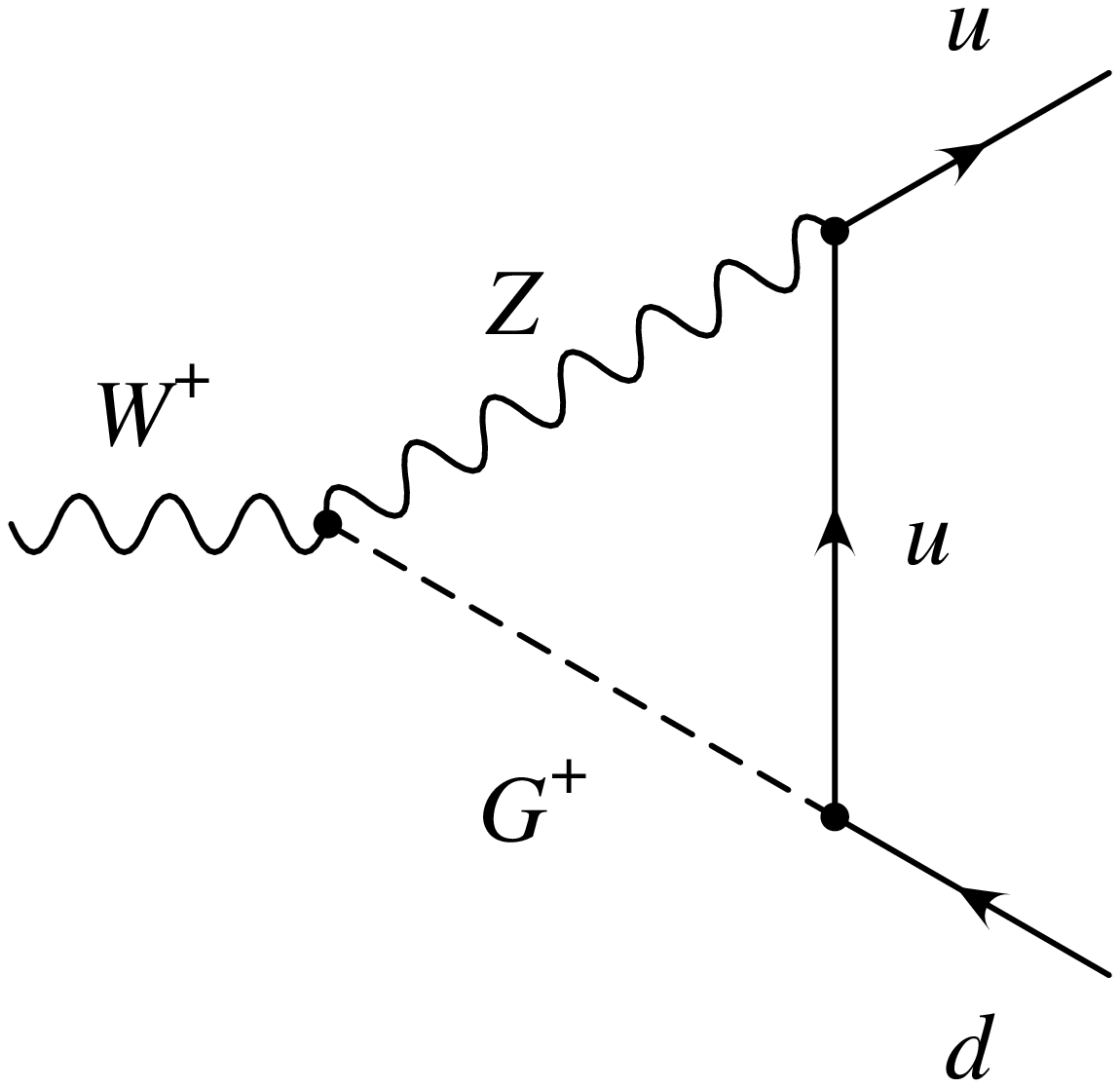,width=3.6cm}
    \epsfig{file=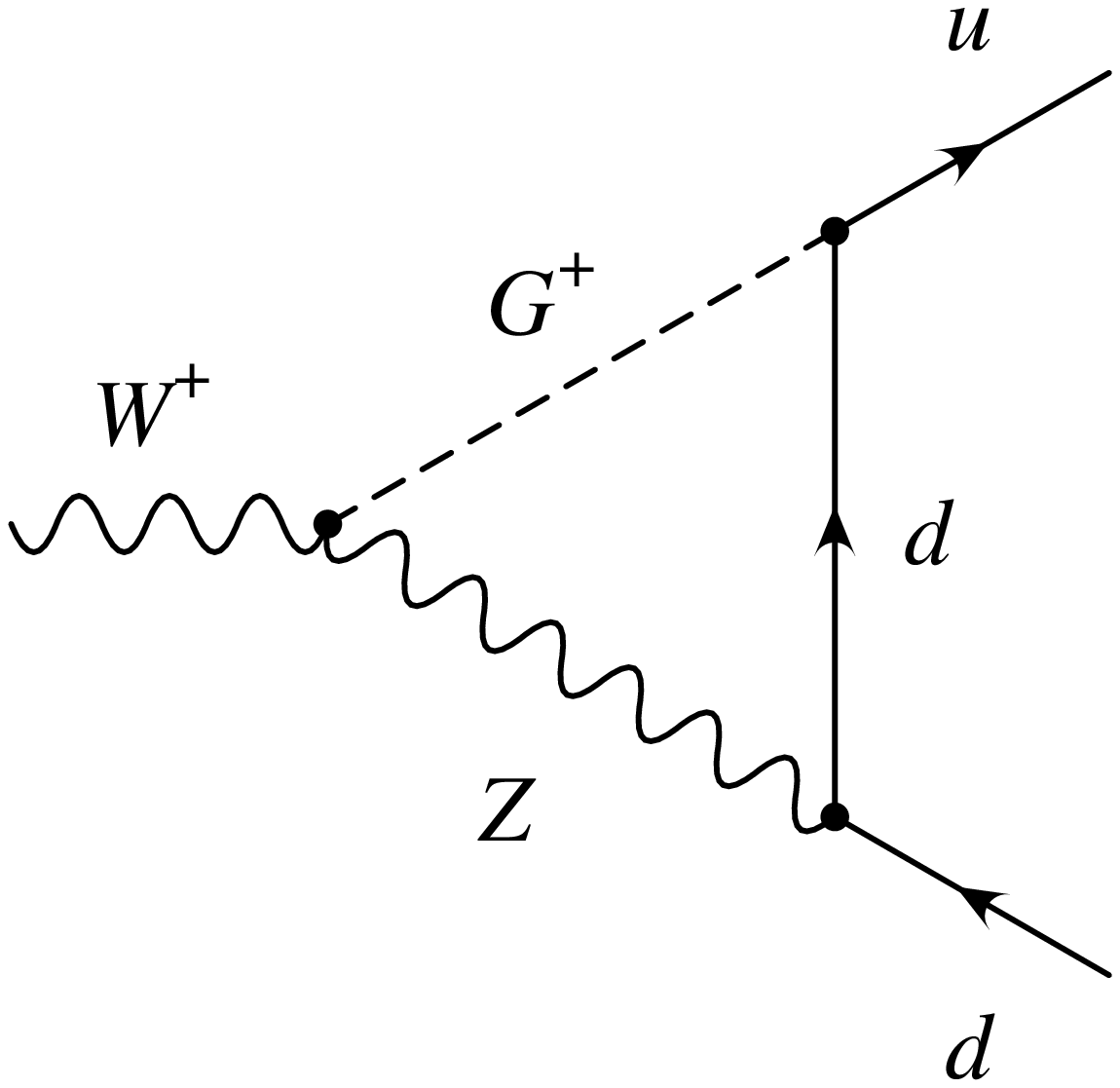,width=3.6cm}\\
    \epsfig{file=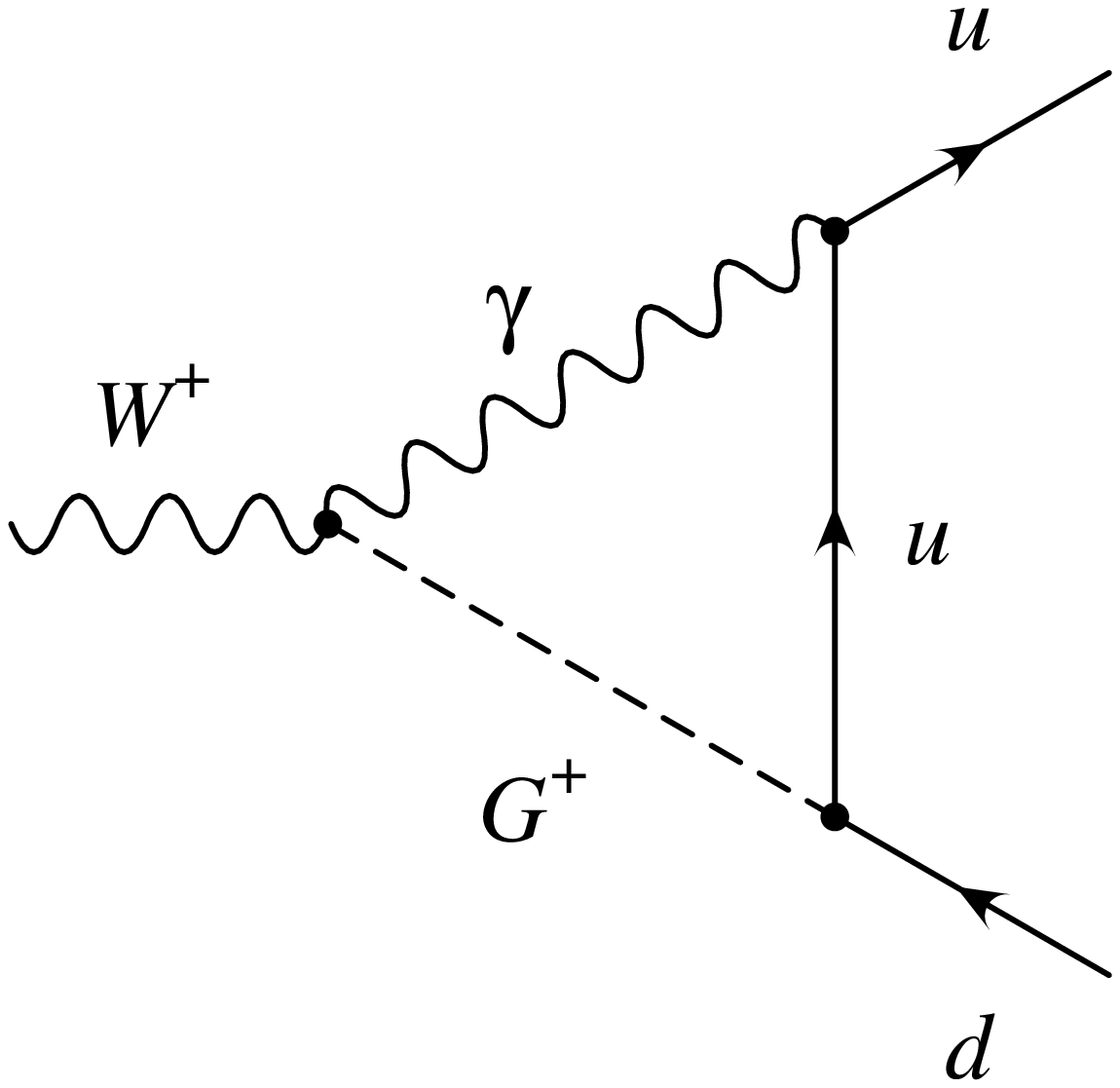,width=3.6cm}
    \epsfig{file=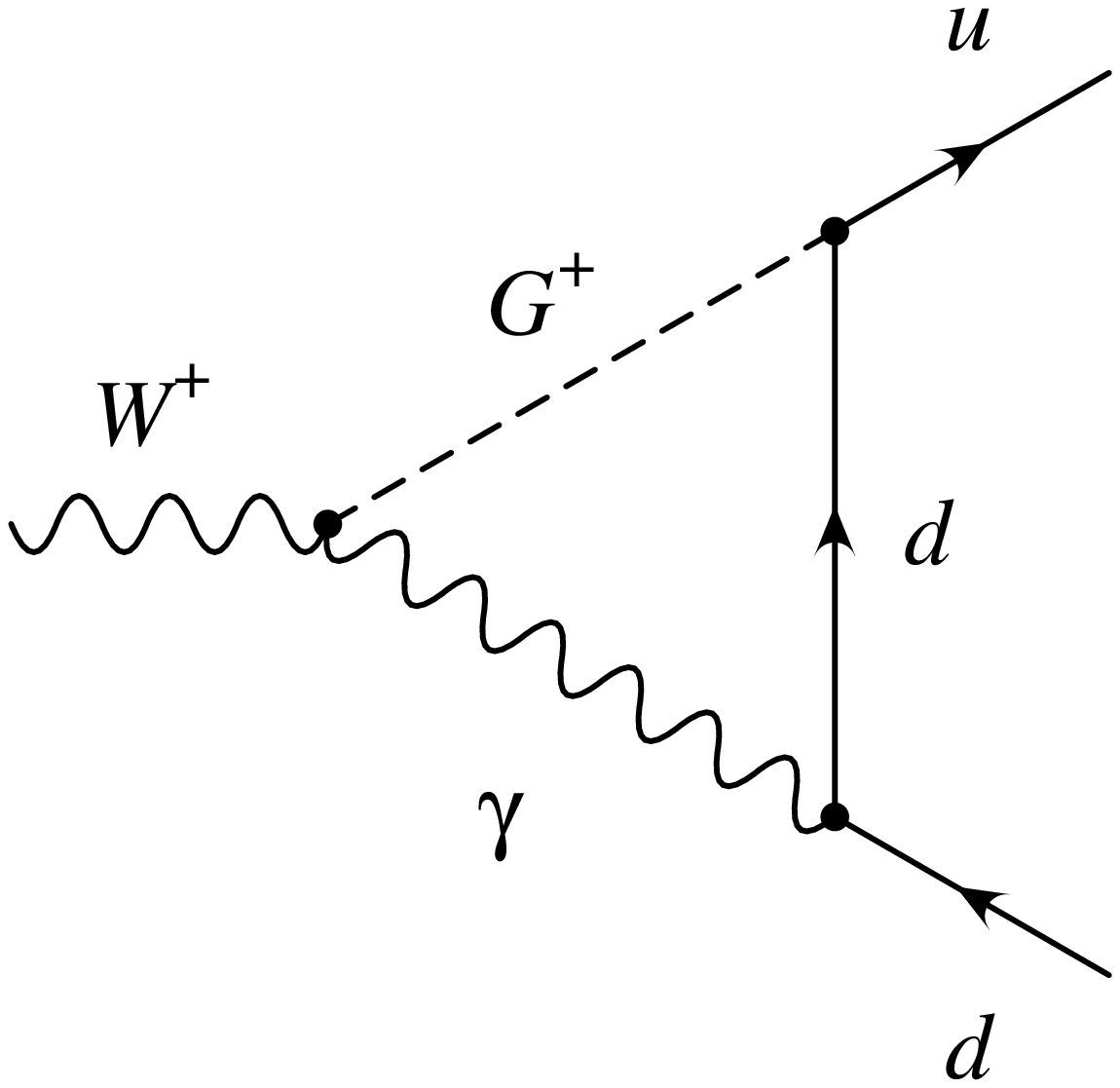,width=3.6cm}\\
   \caption{The diagrams of the electroweak vertex corrections.}
    \label{fig1}
  \end{center}
\end{figure}

In fig.~\ref{fig1} we show the irreducible diagrams that give the
one-loop $W^+ \rightarrow u_I\bar{d}_j$ amplitude. The calculation of
these diagrams using dimensional regularization, is standard. It was
done using the {\em xloops} program \cite{xl}. To keep track
of the divergences it is convenient to introduce the notation
\begin{displaymath}
  \zeta = \frac{2}{D-4} -\gamma_E +\ln 4\pi^2
  -\ln\left(\frac{m_W}{\mu}\right)^2 \enskip ,
\end{displaymath}
where $D$ is the dimension of momentum space ($D \rightarrow 4$),
$\gamma_E$ is the Euler constant and $\mu$ is the arbitrary
renormalization mass.

\begin{figure}[htbp]
  \begin{center}
   \epsfig{file=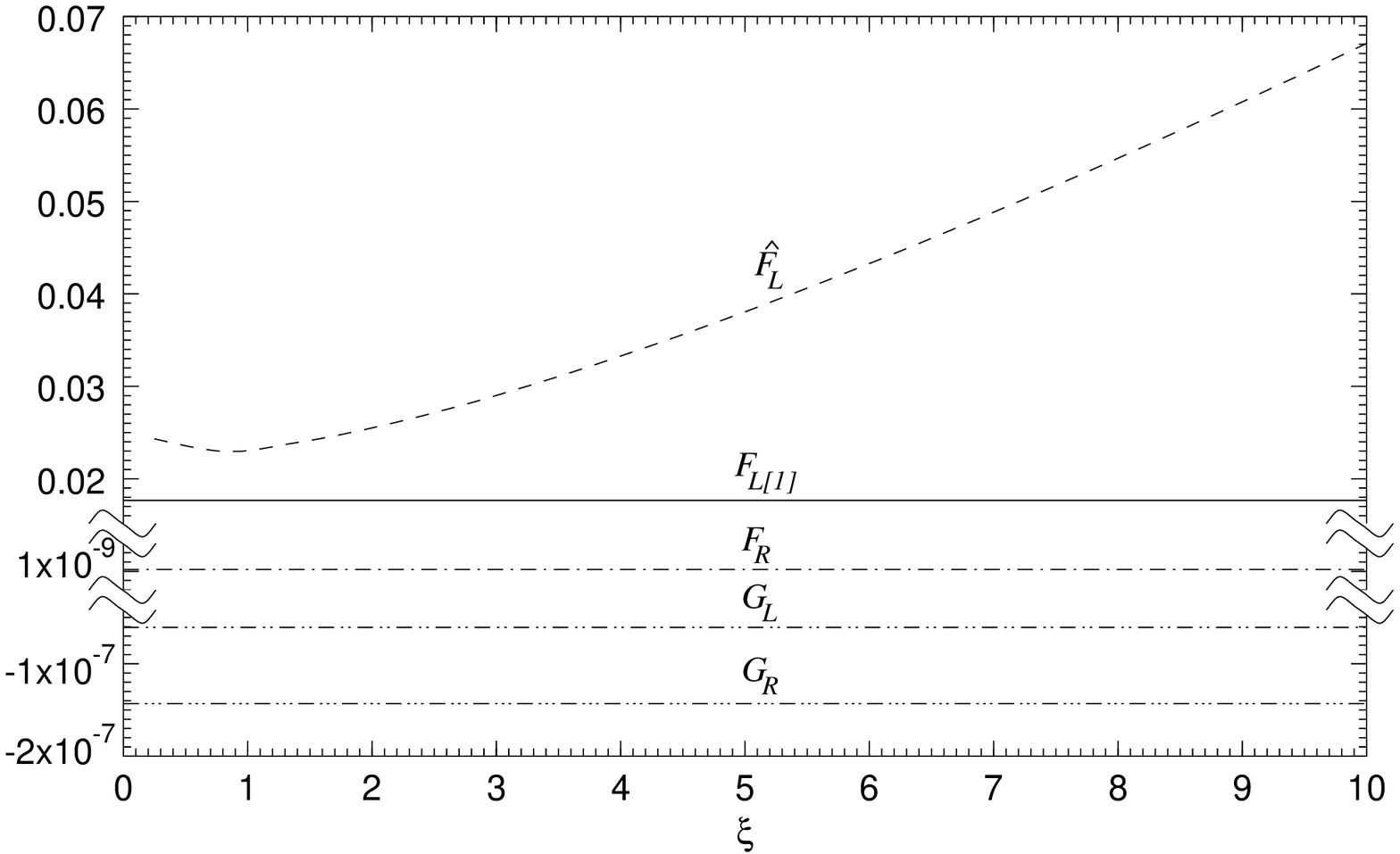,width=14cm}
    \caption{The real part of $F_R$, $G_L$, $G_R$, $\hat{F}_L$ and
      $F_{L\IM}$ for the $W^+\!\rightarrow u\bar{d}$ decay as
      a function of $\xi$.}
    \label{fig2}
  \end{center}
\end{figure}

It is not particularly instructive to show in detail the form
factors. So, we have decided to show explicitly the divergent
contributions and plot the finite parts as a function of $\xi$. In
fig.~\ref{fig2} we display the $\xi$ dependence of the real part of
$F_R$, $G_L$ and $G_R$ for the decay $W^+ \rightarrow u \bar{d}$. As
one can see, these form factors are $\xi$-independent and finite as they
should be. In fact, 
any divergence or gauge dependence here would be
impossible, given the gauge structure of the theory. On
the contrary, $F_L$ is both divergent and $\xi$ dependent, i.e.,
\begin{equation}
  \label{eq7}
  F_L = \frac{e^2}{64\pi^2} \,\zeta\,\left[
  \frac{3\xi+8}{\sin^2\theta_W} + \frac{1}{9\cos^2\theta_W}
   +\frac{m_I^2+m_j^2}{m_W^2}\frac{1}{\sin^2\theta_W}\right]
  \,+\,\hat{F}_L  
\end{equation}
where $\hat{F}_L$ is finite but $\xi$ dependent. This is clearly seen
in fig.~\ref{fig2}. Notice, that the form factors $F_R$, $G_L$ and
$G_R$ are smaller than $\hat{F}_L$ because they are proportional to
the quark masses divided by the $W$ mass. 

\section{The Counter terms}

\subsection{$W$-wave-function renormalization, $\delta Z_W$}

Calculating the $W$-boson self-energy at one-loop and imposing the
on-shell renormalization conditions one obtains \cite{Aok}:
\begin{equation}
  \label{eq8}
  \delta Z_W = \frac{e^2}{96\pi^2\sin^2\theta_W}\,\zeta\, \left[
   22 -3\xi - 2N_g(1+N_c)\right] \,+\,\delta \hat{Z}_W \enskip .
\end{equation}
As before $\delta \hat{Z}_W$ denotes the finite contribution. We will
follow this notation for all counter terms. $N_g = N_c = 3$ are the
number of generations and the number of colors, respectively. We found
that it is convenient to show these parameters explicitly in order to
keep track of the contributions of lepton and quark loops.

From the $W$-self-energy one also obtains the mass counter term, namely
\begin{eqnarray}
  \label{eq9}
  \delta m_W^2 & = & \frac{-e^2}{96\pi^2\sin^2\theta_W}\,\zeta\, \left[
  (34-3\xi) m_W^2 -6m_Z^2 -2 N_g (1+N_c)m_W^2 \phantom{\sum_l}\right. 
  \nonumber \\  
    & & \phantom{ \frac{e^2}{96\pi^2\sin^2\theta_W}\,\zeta\,[} 
   \left.\,+\, 3\sum_l m_l^2
  \,+\,3N_c\sum_{I^\prime j^\prime} |V_{I^\prime j^\prime}|^2
  (m_{I^\prime}^2+m_{j^\prime}^2) \right] \,+\, \delta \hat{m}^2_W
  \enskip .
\end{eqnarray}

\subsection{The coupling counter term, $\delta g$}

It is discussed in great detail in ref.~\cite{Aok} how to obtain
$\delta g$. So, again, we simply summarize our results, which agree
with those in ref.~\cite{Aok} for $\xi = 1$. It is easy to show that
\subnumbers
\begin{equation}
  \label{eq10}
  \frac{\delta g}{g} = \frac{\delta e}{e} - 
                   \frac{\delta \sin\theta_W}{\sin\theta_W} 
\end{equation}
where
\begin{equation}
  \label{eq10b}
  \frac{\delta \sin\theta_W}{\sin\theta_W} = \frac{m_W^2\delta m_Z^2
  -m_Z^2 \delta m_W^2}{2\,m_Z^2\,(m_Z^2-m_W^2)} \enskip .
\end{equation}\normalnumbers
From the $Z$-self-energy one obtains $\delta m_Z^2$. Like the analogue
result shown in eq.~(\ref{eq9}), $\delta m_Z^2$ depends on $\xi$. However,
the combination given by eq.~(\ref{eq10b}) is $\xi$
independent. Furthermore, $\delta e$ is also $\xi$ independent. This
makes the final result
\begin{equation}
  \label{eq11}
  \frac{\delta g}{g} = \frac{e^2}{96\pi^2\sin^2\theta_W}\,\zeta\,
  \left[ N_g(1+N_c) -\frac{43}{2} \right] \,+\,\frac{\delta \hat{g}}{g}  
\end{equation}
fully $\xi$ independent.

\subsection{The quark-fields renormalization}

As it is well known, under renormalization the quark fields are
mixed. Let us write the self-energy of an {\em up}-type quark in the general
form:
\begin{equation}\label{eq13}
  \Sigma_{II^\prime} = \Sigma^L_{II^\prime}(p^2) \slash{p}\gamma_L 
    \, +\, \Sigma^R_{II^\prime}(p^2) \slash{p}\gamma_R
    \, +\, \Sigma^M_{II^\prime}(p^2) [m_I \gamma_L + m_{I^\prime}\gamma_R]
  \enskip .
\end{equation}
Then, using the on-shell renormalization condition one obtains
the matrix elements of the wave function renormalization
constants $\delta Z^L$ \cite{Soa}, namely 
\subnumbers
\begin{equation}\label{eq14a}
  \delta Z^L_{II}  =  -\Sigma^L(m_I^2) - m_I^2 
    \left [ {\Sigma^L}^\prime(m_I^2) +
      {\Sigma^R}^\prime(m_I^2) +
      2{\Sigma^M}^\prime(m_I^2)  \right] \enskip ,
\end{equation}
where ${\Sigma^L}^\prime$ denotes the derivative
$\frac{\partial}{\partial q^2}\Sigma^L$, and for $I^\prime \neq I$
\begin{equation}\label{eq14b}
  {\delta Z^{L}_{II^\prime}}  =  2\frac{
     \left(m_I^2+m_{I^\prime}^2\right) \Sigma^M(m_{I^\prime}^2) +
    m_I m_{I^\prime} \Sigma^R(m_{I^\prime}^2)
   + m_{I^\prime}^2 \Sigma^L(m_{I^\prime}^2)}{m_I^2-m_{I^\prime}^2}
   \enskip .
\end{equation}\normalnumbers
In our case we obtain:
\subnumbers
\begin{equation}
  \label{eq15a}
  \delta Z_{II}^L = \frac{-e^2}{64\pi^2}\zeta \,\left[ \, 
    \frac{1+2\xi}{\sin^2\theta_W} \,+\, \frac{m_I^2 + \sum_{i^\prime}
    |V_{Ii^\prime}|^2 m_{i^\prime}^2}{m_W^2\sin^2\theta_W}
    +\frac{1}{9\cos^2\theta_W}\right] \,+\, \delta \hat{Z}^L_{II}
\end{equation}
for the diagonal terms and 
\begin{equation}
  \label{eq15b}
  \delta Z_{IJ}^L = \frac{-e^2}{32\pi^2\sin^2\theta_W}\zeta 
    \frac{m_I^2 +2m_J^2}{m_J^2 -m_I^2} \sum_{i^\prime}
    V_{Ii^\prime} V_{Ji^\prime}^* \frac{m_{i^\prime}^2}{m_W^2} 
  \,+\, \delta \hat{Z}^L_{IJ} 
\end{equation}\normalnumbers
for the off-diagonal terms. In the latter equation a $\xi$ dependent term in
the divergent part was canceled due to the unitarity of the CKM matrix. The
corresponding result for the {\em down}-type quarks is:
\begin{equation}
  \label{eq16}
  \delta Z_{ij}^L = \frac{-e^2}{32\pi^2\sin^2\theta_W}\zeta 
    \frac{m_i^2 +2m_j^2}{m_j^2 -m_i^2} \sum_{I^\prime}
    V_{I^\prime i} V_{I^\prime j}^* \frac{m_{I^\prime}^2}{m_W^2} 
  \,+\, \delta \hat{Z}^L_{ij} 
\end{equation}
and the diagonal part is identical to eq.~(\ref{eq15a}) replacing $I$ by
$i$ and $I^\prime$ by $i^\prime$. It is interesting to point out that
the matrices $\delta Z^L$ are neither hermitian nor anti-hermitian. Of
course, they can be decomposed in a sum of such matrices, $\delta Z^L
= \delta Z^{LH} + \delta Z^{LA}$. However, one should realize that the
divergence is present both in $\delta Z^{LH}$ and in $\delta
Z^{LA}$. In fact, from eq.~(\ref{eq15b}) it is straightforward to obtain
\subnumbers
\begin{equation}\label{eq17a}
  \delta Z_{IJ}^{LH} = \frac{-e^2}{64\pi^2\sin^2\theta_W}\zeta 
    \sum_{i^\prime} V_{Ii^\prime} V_{Ji^\prime}^* 
   \frac{m_{i^\prime}^2}{m_W^2} \,+\, finite 
\end{equation}
and
\begin{equation}\label{eq17b}
  \delta Z_{IJ}^{LA} = \frac{-3 e^2}{64\pi^2\sin^2\theta_W}\zeta 
    \frac{m_I^2 +m_J^2}{m_J^2 -m_I^2} \sum_{i^\prime}
    V_{Ii^\prime} V_{Ji^\prime}^* \frac{m_{i^\prime}^2}{m_W^2} 
  \,+\, finite \enskip . 
\end{equation}\normalnumbers
Clearly, eq.~(\ref{eq15a}) shows that the diagonal terms of $\delta Z^L$
are real. These remarks will be important in paragraph 5, when we
consider the renormalization of the CKM matrix.

\section{The $W^+$ decay into leptons}

Using eqs.~(\ref{eq7}),(\ref{eq8}),(\ref{eq11}),(\ref{eq15a},b) and (\ref{eq16})
it is easy to obtain:
\begin{eqnarray}
  \label{eq18}
\lefteqn{  F_L + \frac{\delta g}{g} + \halb\delta Z_W +\halb\delta Z_{II}^{L*}
  + \halb\delta Z_{jj}^L}\\
  & = & \frac{e^2}{128\pi^2\sin^2\theta_W} \zeta \,
  \frac{m_I^2-\sum_{i^\prime}|V_{Ii^\prime}|^2 m_{i^\prime}^2
      + m_j^2-\sum_{I^\prime}|V_{I^\prime j}|^2 m_{I^\prime}^2 }{m_W^2}
     + finite \enskip .\nonumber
\end{eqnarray}
Notice that there are no divergences proportional to the gauge
parameter $\xi$. If $V_{Ii^\prime}=\delta_{Ii^\prime}$ and 
$V_{I^\prime j}=\delta_{I^\prime j}$, i.e., if the CKM matrix is the
unit matrix, the divergent term is identically zero. In this case, we
call the above combination of $F_L$ and counter terms
$F_{L\IM}$.\footnote{Obviously in $F_{L\IM}$ $\delta g$ and $\delta
  Z_W$ are not calculated with a unit CKM matrix.}

From eq.~(\ref{eq4}) it is now clear that the one-loop leptonic decay
amplitude $W^+ \rightarrow l^+ \nu_l$ can be written as
\begin{equation}
  \label{eq19}
  T_1^l = A_L F_{L\IM} + B_R G_R \enskip ,
\end{equation}
where in $F_{L\IM}$ and $G_R$ the leptonic masses are used
and in eq.~(\ref{eq2}) and (\ref{eq3}) we set $N_c=1$. The form
factors $F_R$ and $G_L$ are proportional to $m_I$. Hence, they vanish
for massless neutrinos.
As we have shown $T_1^l$ is finite, as it should be. Furthermore,
fig.~\ref{fig2}, where we show $F_{L\IM}$ as a function of
$\xi$, clearly proves that the one-loop leptonic amplitude $T_1^l$ is
also gauge independent. Having established the finiteness and the gauge
independence of $F_{L\IM}$ we are now in a position to return
to eq.~(\ref{eq4}) and consider the $\delta V_{Ij}$ counter term.

\section{The CKM counter term}

Let us consider the $W$-quark coupling in the standard model
Lagrangian. Introducing an obvious matrix notation we write 
\begin{equation}
  \label{eq20}
  {\cal L} = -\frac{g}{\sqrt{2}} \bar{U}_L V D_L W_\mu + h.c. \enskip ,
\end{equation}
where $U_L$ and $D_L$ are the left-handed up and down
quark fields respectively. Leaving aside the renormalization of $g$
and of the $W$-field, let us focus our attention in the
renormalization of the quark fields and $V$. In the former work of DS
the matrix $V$ is multiplicatively renormalized, i.e.
\begin{equation}
  \label{eq21}
  V \rightarrow U_1 V U_2 = V +\delta U_1 V + V\delta U_2 \enskip ,
\end{equation}
where $U_1$ and $U_2$ are unitary matrices. Then, introducing the
usual quark wave-function renormalization,
\begin{eqnarray*}
  \bar{U}_L & \rightarrow & \bar{U}_L {Z^{1/2}_{UL}}^\dagger \\
  D_L & \rightarrow & Z^{1/2}_{DL} D_L 
\end{eqnarray*}
eq.~(\ref{eq20}) becomes:
\begin{eqnarray}
  \bar{U}_L V D_L & \rightarrow & \bar{U}_L {Z^{1/2}_{UL}}^\dagger 
        U_1 V U_2 Z^{1/2}_{DL} D_L \nonumber \\
  \label{eq23}
    & = & \bar{U}_L \left\lbrace \,V 
     \,+\, \viertel \left[ \delta Z_{UL}^\dagger + \delta Z_{UL}\right] V
     \,+\, \viertel V \left[ \delta Z_{DL} + \delta Z_{DL}^\dagger\right] 
    \right. \\ & & \phantom{\left[ \delta Z_{UL}^\dagger \right]}
     \,+\, \viertel \left[ \delta Z_{UL}^\dagger - \delta Z_{UL}\right] V
     \,+\, \viertel V \left[ \delta Z_{DL} - \delta Z_{DL}^\dagger\right] 
    \nonumber \\ & & \left.\phantom{\left[ \delta Z_{UL}^\dagger \right]}
     \,+\, \delta U_1V \,+\, V\delta U_2 \,\right\rbrace\,D_L \nonumber
     \enskip ,
\end{eqnarray}
where, for convenience, we have split the $\delta Z$ matrices into its
hermitian and anti-hermitian parts. Because the unitarity of
the matrices $U_i$ implies that the $\delta U_i$ are anti-hermitian,
DS concluded that $\delta V =\delta U_1 V + V\delta U_2$ is required to
absorb the divergence in the anti-hermitian parts of $\delta
Z_L$. Hence, they have introduced the following renormalization
condition: 
\begin{equation}
  \label{eq24}
  \delta V = -\viertel \left[ \delta Z_{UL}^\dagger - \delta Z_{UL}\right] V
     \,-\, \viertel V \left[ \delta Z_{DL} - \delta Z_{DL}^\dagger\right] 
   \enskip .
\end{equation}
Of course, there are still divergences in the hermitian part of
$\delta Z$, but, as we will see, they are the ones needed to cancel the
divergences in the vertex contribution to $F_L$. In fact, using
eqs.~(\ref{eq17a},b) and (\ref{eq15a},b) it is straightforward to obtain:
\begin{eqnarray}
  \label{eq25}
  \halb \left[\delta Z_{UL}^H V + V \delta Z_{DL}^H\,\right]_{Ij} 
   & = & \frac{-e^2}{128\pi^2\sin^2\theta_W} \zeta
   \,\left[\sum_{i^\prime J} V_{Ii^\prime} V_{Ji^\prime} V_{Jj}
   \frac{m_{i^\prime}^2}{m_W^2} \,+\,  \frac{m_{I}^2}{m_W^2}V_{Ij} 
 \right.\nonumber \\ & & \left. \qquad
   \,+\, \sum_{I^\prime i^\prime} V_{I i^\prime} V_{I^\prime i^\prime} V_{I^\prime j}
   \frac{m_{I^\prime}^2}{m_W^2} \,+\,  \frac{m_{j}^2}{m_W^2}V_{Ij}
   \,\right]\;+\;finite \enskip . 
\end{eqnarray}
In the equation above, when using the diagonal elements of the matrix
$\delta Z^H$, only the contribution of the second term of
eq.~(\ref{eq15a}) is explicitly shown.
The other two terms are
irrelevant for the discussion since they cancel with similar divergences
coming from $F_L$, $\delta Z_W$ and $\delta g$.

Now, the unitarity of $V$ reduces eq.~(\ref{eq25}) to the form:
\begin{equation}
  \label{eq26}
  \halb \left[\delta Z_{UL}^H V + V \delta Z_{DL}^H\,\right]_{Ij} 
 = \frac{-e^2}{64\pi^2\sin^2\theta_W} \zeta \,
    \frac{m_I^2+m_j^2}{m_W^2} \,V_{Ij} \enskip ,
\end{equation}
which is exactly what we need to cancel a similar divergence in
$V_{Ij} F_L$, namely the third term in eq.~(\ref{eq7}).

Hence, from the point of view of canceling the divergences in $T_1$,
the renormalization proposal by DS works. In other words, it is
sufficient to choose $\delta V$ as the divergent part of the right
hand side of eq.~(\ref{eq24}) to obtain a finite one-loop
amplitude. DS have also included in $\delta V$ the finite
contributions stemming from $\delta Z^A$. We have checked that this
gives rise to a gauge dependent result. 

To solve this problem let us define the quantity
\begin{eqnarray}
  \label{eq27}
  \delta X_{ud} & = & \halb V_{ud}\left[ 
    \delta Z_{uu}^{L*} -\delta Z_{uu \IM}^{L*} 
   +\delta Z_{dd}^{L} -\delta Z_{dd \IM}^{L} \right]
    \nonumber \\ & &
   +\halb \sum_{I^\prime\neq u} \delta Z^{L*}_{I^\prime u} V_{I^\prime d}
   +\halb \sum_{j^\prime\neq d} V_{uj^\prime} \delta Z^{L}_{j^\prime d}
   \enskip ,
\end{eqnarray}
which obviously represents the difference between the
``leptonic''\footnote{Here leptonic means that no mixing takes place
  among the different generations. Of course, for calculating the
  $\delta Z_{\IM}$ renormalization constants, massive quarks
  were used.} 
and the quark transition amplitude. Notice
that $\delta Z_{uu\IM}^{L*}$ is given by eq.~(\ref{eq15a})
but replacing the CKM by the unit matrix.  
After introducing the quantity $\delta X_{Ij}$ it is clear that
eq.~(\ref{eq4}) can be rewritten as
\begin{equation}
  \label{eq28}
  T_1 = V_{Ij} \left[ A_L F_{L\IM} + A_R F_R + B_L G_L +
  B_R G_R \right] + A_L\left[\delta X_{Ij} +\delta V_{Ij}\right]\enskip .
\end{equation}
Having proved that the first term of eq.~(\ref{eq28}),
proportional to $V_{Ij}$, is both finite and gauge independent, it is
obvious that the CKM counter term should be 
\begin{equation}
  \label{eq29}
  \delta V_{Ij} = - \delta X_{Ij} \enskip .
\end{equation}
This is our main result. On physical terms what we are saying is that
all contributions to the $T_1$ amplitude arising from the
renormalization of the quark mixing are canceled by the CKM
counter term. This $\delta V$ is an alternative to the one proposed by
GGM which requires the use of quark wave function renormalization
constants at zero momentum. Both schemes lead to gauge invariant
results. In fact, the unitarity of the CKM matrix implies that $\delta
X$ is gauge independent.

\section{Conclusions}

Beyond tree-level, quarks with the same electric charge get mixed
under renormalization. Then, the amplitude for the $W^+ \rightarrow u
\bar{d}$ explicitly depends on these flavor-changing renormalization
constants. Therefore, to obtain a finite amplitude it is essential to
renormalize the corresponding element of the CKM matrix, $V$. Using
the on-shell renormalization scheme and the $R_\xi$ gauge we have
shown how to construct the CKM counter term matrix $\delta V$. Our
final result is given in eq.~(\ref{eq29}). With this prescription the
tree-level relation 
\begin{displaymath}
  T(W^+\rightarrow u \bar{d}) = V_{ud} N_c T(W^+\rightarrow e^+ \nu_e)
  \enskip ,
\end{displaymath}
living aside $\alpha_s$ corrections and obvious kinematic
differences, is maintained at the next order. We have proved that at
one-loop one obtains a finite and gauge independent amplitude. 
It is interesting to point out that a finite amplitude $T_1$ can only be
obtained, if the CKM matrix is unitary. This is particular important
in view of some recent discussions about a possible non-unitarity of
this matrix \cite{NU}.

\section{Acknowledgement}

We thank Paolo Gambino for pointing out an error in a previous version
of this work. This work is supported
by Fundaç\~ao para a Ci\^encia e Tecnologia under contract
No. CERN/P/FIS/15183/99. L.B. is supported by JNICT
under contract No. BPD.16372.

\begin{appendix}

\end{appendix}





\end{document}